\documentclass[%
 reprint,
 amsmath,amssymb,
 aps,
]{revtex4-2}

\usepackage{graphicx}
\usepackage{dcolumn}
\usepackage{bm}
\usepackage{xcolor}

\allowdisplaybreaks

\begin{document}

\preprint{APS/123-QED}

\title{Cabello's nonlocality argument for multisetting high-dimensional systems and its experimental test}

\author{Ming Yang$^1$}
\author{Dongkai Zhang$^{1,2}$}
\email{zhangdk@hqu.edu.cn}
\author{Lixiang Chen$^1$}
\email{chenlx@xmu.edu.cn}
\affiliation{$^1$Department of Physics, Xiamen University, Xiamen 361005, China}
\affiliation{$^2$College of Information Science and Engineering, Fujian Provincial Key Laboratory of Light Propagation and Transformation, Huaqiao University, Xiamen 361021, China}

\date{\today}

\begin{abstract}
Recent advancements have expanded Hardy's nonlocality arguments into multisetting and multidimensional systems to enhance quantum correlations. In comparison with Hardy's nonlocality argument, Cabello's nonlocality argument (CNA) emerges as a superior choice for illustrating nonlocal features. An open question persists regarding the potential extension of CNA to arbitrary ($k$, $d$) scenarios. Here, we answer this question both in theory and experiment.  Theoretically, by utilizing compatibility graphs, we construct a new logical framework for multisetting and multidimensional CNA, demonstrating an increase in the maximum successful probability with setting $k$ and dimension $d$. Experimentally, by employing controllable photonic orbital angular momentum entanglement, we exhibit nonlocality with an experimentally recorded probability of 20.29\% in the (2, 4) scenario and 28.72\% in the (6, 2) scenario. Our work showcases a sharper contradiction between quantum mechanics and classical theory, surpassing the bound limited by the original version. 
\end{abstract}

\maketitle


\section{\label{sec:level1}Introduction}

The exploration of quantum nonlocality traces its origins back to the famous EPR paradox of 1935, where the foundational concept of local realism was introduced \cite{epr}. A decisive development occurred in 1964 with Bell inequality, which decisively refuted the EPR paradox by revealing irreconcilable predictions between quantum mechanics and local realism, thereby establishing the concept of ``quantum nonlocality" \cite{Bell1965}. Since then, this domain has been extensively scrutinized, yielding myriad demonstrations of quantum nonlocality. Among these demonstrations, the subclass of ``inequality-free" proofs has garnered significant attention within the physics community. The GHZ paradox, an early exemplar of an ``inequality-free" proof, established inevitable violations of local realism for a three-body quantum entangled state, namely, the GHZ state \cite{Greenberger1990}. In 1990s, Hardy advanced a comparable ``inequality-free" proof applicable to two-body entanglement states \cite{Hardy1992,Hardy1993}, termed Hardy’s paradox or Hardy's nonlocality argument (HNA). Although effective for all purely entangled bipartite qubit states, excluding maximally entangled states, HNA was constrained to a maximum successful probability of approximately 9\%. In 2002, Cabello introduced an alternative nonlocality argument, initially focusing on three-body quantum entangled states \cite{Cabello2002}, which was later applied to two-body qubit entangled states \cite{Liang2005}. In direct comparison with HNA, Cabello's nonlocality argument (CNA) emerged as a more potent tool, capable of attaining a maximum successful probability of approximately 11\% \cite{Kunkri2006}. Moreover, a comprehensive comparison between HNA and CNA, grounded in fundamental principles, revealed the superior discriminatory power of CNA in detecting post-quantum no-signaling correlations in bipartite systems \cite{zoka2016local}. Beyond theoretical advancements, the pragmatic implications of CNA in the realm of quantum information have been acknowledged, notably in the development of device-independent random number expansion protocols by Li and co-workers \cite{Li2015}. Their findings underscored the practical superiority of CNA over HNA in generating randomness, accentuating the utilitarian value of CNA. 

Recent years have witnessed a growing interest in extending nonlocality arguments to achieve more robust quantum correlations. Both Bell inequality \cite{Collins2002,Zohren2008,Dada2011,Tavakoli2016,Salavrakos2017} and HNA \cite{Boschi1997,Chen2013,Chen2017,Meng2018,Zhang2020} have been extended to multi-setting high-dimensional versions, highlighting a sharper contradiction between quantum mechanics and classical theory. we note that, the aforementioned investigations of CNA were limited to a 2-setting 2-dimensional scenario. This raises the question: being compared with Bell inequality and HNA, whether the maximum successful probability of CNA increase with both setting and dimension in arbitrary ($k$, $d$) scenarios? Such an extension is of paramount physical importance as it effectively demonstrates the completeness of quantum mechanics \cite{Colbeck2008,Stuart2012}. Furthermore, as entanglement tends to become more resilient to noise with increasing dimension, this extension facilitates stronger tests of quantum nonlocality \cite{Ecker2019,Zhu2021,Gois2023,Qu2022}, culminating in device independence for high-dimensional systems \cite{Tavakoli2021,Sarkar2021}. Here, we present a multi-setting high-dimensional version of CNA. In theory, we explore a compatibility graph to demonstrate this new version of CNA, revealing that the maximum successful probability increases with setting $k$ and dimension $d$. Notably, our numerical findings demonstrate an expanding gap in maximum successful probability between CNA and HNA as the dimension increases. In experiment, we demonstrate the CNA with high-dimensional biphoton orbital angular momentum (OAM) entanglement states. Crucially, we showcase the nonlocality with an experimentally recorded probability of 20.29\% in the (2, 4) scenario and of 28.72\% in the (6, 2) scenario, thereby generalizing the CNA \cite{Kunkri2006} to a truly multi-setting high-dimensional scenario. Our findings show that nonlocal events can indeed increase with the dimension of the system, significantly surpassing the Device-independent bounds limited by the original version \cite{Rai2021}, which established a constant upper bound for CNA regardless of dimension.

\section{Formalism for Cabello-type paradox}

Let us consider a general scenario where each of Alice and Bob can choose $k$ sets
of measurements, and each set of measurements can get $d$ outcomes. To facilitate the description of the argument, we denote Alice's measurements as $M_{2i-1}$ and Bob's as $M_{2i}$, corresponding to von Neumann measurements $\left| {{M_{2i - 1,s}}} \right\rangle \left\langle {{M_{2i - 1,s}}} \right|$ and $\left| {{M_{2i,s}}} \right\rangle \left\langle {{M_{2i,s}}} \right|$, where $i \in \left\{ {1,2, \ldots ,k} \right\}$ and $s\in\left\{ {1,2, \ldots ,d} \right\}$. Then an extended version of CNA in general ($k, d$) scenario can be presented as the following chain of probabilities hold:
\begin{equation}
\label{eq:1}
    P\left( {{M_1} > {M_{2k}}} \right) = {P_1},
\end{equation}
\begin{equation}
\label{eq:2}
    P\left( {{M_J} > {M_{J + 1}}} \right) = {P_2},
\end{equation}
\begin{equation}
\label{eq:3}
    P\left( {{M_i} > {M_{i + 1}}} \right) = 0,{\rm{ for }}\ i = 1,2, \ldots, 2k-1,\ i \ne J,
\end{equation}
Within any local hidden variable theory, following Eqs.~(\ref{eq:1}),~(\ref{eq:2}), and~(\ref{eq:3}), we straightforwardly obtain Cabello's fraction $ {P_1} - {P_2} \le 0$. Conversely, quantum mechanics allows the suitable choice of measurements to satisfy Eqs.~(\ref{eq:1}),~(\ref{eq:2}), and~(\ref{eq:3}), yet 
\begin{equation}
\label{eq:4}
    {P_1} - {P_2} > 0.
\end{equation}
Note that $J$ can be chosen as any positive integer less than $2k$, rather than $J=2$ in the original CNA formulation \cite{Kunkri2006}. Thus, we represent scenarios involving three parameters $k$, $d$, $J$ as $(k, d, J)$ for clarity.

\begin{figure}[b]
\includegraphics[width=0.3\textwidth]{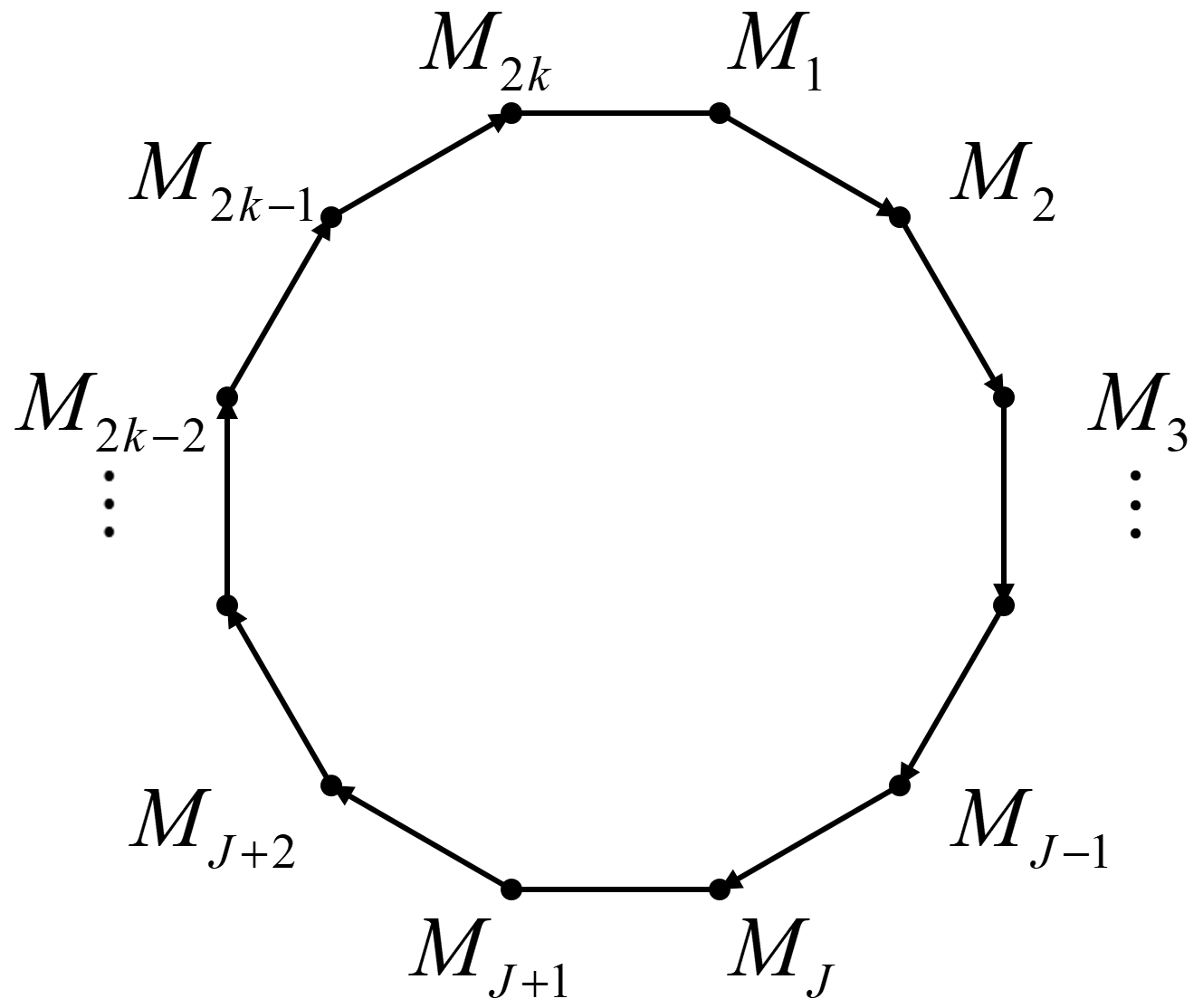}
\caption{\label{fig:cycle} A cycle of new compatibility graph when Eq.~(\ref{eq:3}) holds. Edge with arrow from $M$ to $M'$ denote ordered pair $(M,M')$, implies $M$ can reach $M'$ without meaning $M'$ can reach $M$. When $(M_{2k},M_i)$ and $(M_J,M_{J+1})$ hold, the cycle forms a directed cycle.}
\end{figure}

Recent insights into quantum nonlocality and contextuality have leveraged graph theory methods \cite{Cabello2014,Sohbi2019,Tang2022,Abramsky2012}, such as Cabello's exclusivity graph, to characterize quantum correlations\cite{Cabello2014}. This framework facilitated the identification of experimental scenarios capable of generating correlations on demand by selecting graphs possessing specific properties, while also categorizing quantum correlations based on the analysis of their graph characteristics. Here we introduce the compatibility graph to prove the Cabello-type paradox. Within this graph, vertices symbolize measurements, while edges signify pairwise compatibility relations, indicating that connected measurements can be jointly measured \cite{Budroni2022}. According to Eqs.~(\ref{eq:1}),~(\ref{eq:2}), and~(\ref{eq:3}), we depict the measurements ${{M_1},{M_2}, \ldots ,{M_{2k}}}$ as a cycle \cite{Bender2010}, as shown in Fig.~\ref{fig:cycle}. Now, we introduce a new structure into the compatibility graph: If two adjacent vertices $M$ and $M'$ on the graph satisfy $M \le M'$, i.e., the outcome of $M$ less than or equal to that of $M'$, then we construct an ordered pair $\left( {M,M'} \right)$ \cite{Wolf1998} (see Fig.~\ref{fig:cycle}).
Under this new structure, we formulate the following theorem:\par
\textit{Theorem 1}. If the compatibility graph constructed from a group of measurements contains a directed cycle \cite{Bender2010}, then under local realism, the values of all measurements represented by the vertices on the directed cycle are equal.\par
\textit{Proof}. Suppose there exists a directed cycle $\left( {{M_1},{M_2}, \ldots ,{M_K},{M_1}} \right)$, on which any two vertices ${M_i}$ and ${M_j}$ ($i < j$) are mutually reachable\cite{Schmidt2010}. Specifically, ${M_i}$ can reach ${M_j}$ via a sequence of vertices $({M_i},{M_{i + 1}}, \ldots ,{M_j})$, and ${M_j}$ can reach ${M_i}$ via $({M_j},{M_{j + 1}}, \ldots ,{M_K},{M_1}, \ldots ,{M_{i - 1}},{M_i})$. As the ordered pair $\left( {M,M'} \right)$ represents transitive relation $M \le M'$ \cite{Simovici2008}, these sequences indicate ${M_i} \leq {M_{i + 1}} \leq \ldots \leq {M_j}$ and ${M_j} \leq {M_{j + 1}} \leq \ldots \leq {M_{i - 1}} \leq {M_i}$, meaning that both ${M_i} \le {M_j}$ and ${M_j} \le {M_i}$ hold. Consequently, any two vertices satisfy ${M_i} = {M_j}$, and thus all values represented by vertices on the directed cycle are equal.

Through \textit{Theorem 1}, we can systematically investigate various paradoxes, such as Hardy's paradox and the Cabello-type paradox. Now, we give the explanation of Eqs.~(\ref{eq:1})-(\ref{eq:4}) from a probabilistic perspective. Assuming Eq.~(\ref{eq:3}) holds and events ${M_{2k}} < {M_1}$ and ${M_J} \le {M_{J + 1}}$ occur, ordered pairs $\left( {{M_{2k}},{M_1}} \right)$ and $\left( {{M_J},{M_{J + 1}}} \right)$ are constructed, forming a directed cycle. Since ${M_{2k}}$ is strictly less than ${M_1}$, the values of measurements are unequal. By Theorem 1, under local realism, such an event ${M_{2k}} < {M_1} \cap {M_J} \le {M_{J + 1}}$ is impossible. Employing the addition rule in probability theory \cite{Shiryaev2016}, we deduce
\begin{equation}
    \begin{array}{l}
{P_1} + (1 - {P_2})\\
 = P({M_{2k}} < {M_1}) + P({M_J} \le {M_{J + 1}})\\
 = P({M_{2k}} < {M_1} \cup {M_J} \le {M_{J + 1}})\\
 + P({M_{2k}} < {M_1} \cap {M_J} \le {M_{J + 1}})\\
 = P({M_{2k}} < {M_1} \cup {M_J} \le {M_{J + 1}}) \le 1,
\end{array}
\end{equation}
implying that ${P_1} - {P_2} \le 0$, which completes the proof. In the case of $(2, 2, 2)$ scenario, the paradox is equivalent to the original CNA \cite{Kunkri2006} (see Appendix A for more details). Obviously, with a suitable choice of measurements, quantum mechanics allows ${P_1} - {P_2} > 0$. In the subsequent theoretical and experimental procedures, we focus on testing the Cabello's logical structure. Theoretically, we employ numerical simulations to determine the maximum Cabello's fraction and the optimal measurement states. Experimentally, we perform the detection of all the desired measurement states to illustrate a sharper contradiction between quantum mechanics and classical theories.

We first establish the strategy for finding the maximum Cabello's fraction, i.e., the maximum value of the nonlocal probability ${P_1} - {P_2}$. In any $(k,d,J)$ scenario, each set of measurements, such as $M_i$, can be represented as a \(SU(d)\) unitary matrix, with the rows of the matrix representing measurement bases \cite{Zhang2020}. For numerical calculation convenience, we set \(M_1\) and \(M_{2k}\) as the identity matrix, i.e., both $\left\{ {|{M_{1,1}}\rangle,|{M_{1,2}}\rangle, \ldots |{M_{1,d}}}\rangle \right\}$ and $\left\{ |{{M_{2k,1}}\rangle,|{M_{2k,2}}\rangle, \ldots |{M_{2k,d}}}\rangle \right\}$ are standard orthogonal bases, and reformulated two-qudit entangled state $\left| \varphi \right\rangle = \sum\nolimits_{a,b = 1}^d {{h_{ab}}{{\left| a \right\rangle }_A} \otimes {{\left| b \right\rangle }_B}}$ as a matrix $H = {\left( {{h_{ab}}} \right)_{1 \le a,b \le d}}[16]$. Consequently, we calculate the probability 
\begin{eqnarray}
\label{eq:5}
    \begin{gathered}
  P\left( {{M_i} > {M_j}} \right) = \sum\limits_{s > t} {{{\left| {\left\langle \varphi  \right.|\left. {{M_{i,s}}} \right\rangle \left.| {{M_{j,t}}} \right\rangle } \right|}^2}}  \hfill \\
   = \left\{ {\begin{array}{*{20}{c}}
  {\sum\limits_{s > t} {{{| {\sum\limits_{a,b} {m_{i,s,a}^*m_{j,t,b}^*{h_{ab}}} } |}^2} \quad {\text{if }}i{\text{ is odd}}} } \\ 
  {\sum\limits_{s > t} {{{| {\sum\limits_{a,b} {m_{i,s,b}^*m_{j,t,a}^*{h_{ab}}} } |}^2} \quad {\text{if }}i{\text{ is even}}} } 
\end{array}}. \right. \hfill \\ 
\end{gathered} 
\end{eqnarray}
Note that $i$ and $j$ have different parity, i.e., when $i$ is odd, $j$ is even; when $i$ is even, $j$ is odd. Here, \(|M_{i,s} \rangle = \sum_{a=1}^d m_{i,s,a} |a\rangle\) and \(|M_{j,t} \rangle = \sum_{b=1}^d m_{j,t,b} |b\rangle\) when $i$ is odd, and \(|M_{i,s} \rangle = \sum_{b=1}^d m_{i,s,b} |b\rangle\) and \(|M_{j,t} \rangle = \sum_{a=1}^d m_{j,t,a} |a\rangle\) when $i$ is even. Then, from Eq. (1), we calculate ${P_1} = {\sum\nolimits_{a > b} {\left| {{h_{ab}}} \right|} ^2}$; from Eq. (2), we calculate ${P_2}={\sum\nolimits_{s > t} {{{| {\sum\nolimits_{a,b} {m_{J,s,a}^*m_{J + 1,t,b}^*{h_{ab}}} } |}^2}}}$ when $J$ is odd, and ${P_2} ={\sum\nolimits_{s > t} {{{| {\sum\nolimits_{a,b} {m_{J,s,b}^*m_{J + 1,t,a}^*{h_{ab}}} } |}^2}}}$ when $J$ is even. From Eq. (3), we calculate $P(M_i > M_{i+1}) =\sum\nolimits_{s > t}{{{\left| \langle \varphi |M_{i,s} \rangle | M_{i + 1,t}\rangle \right|}^2}}=0$, which implies the orthogonal relations \(|M_{i+1,t} \rangle \perp \langle M_{i,s} |\varphi \rangle\) for all $s>t$ . As a result, we can uniquely determine measurements $\left\{ {|{M_{i+1,1}}\rangle,|{M_{i+1,2}}\rangle, \ldots |{M_{i+1,d}}}\rangle \right\}$ (or $\left\{ {|{M_{i,1}}\rangle,|{M_{i,2}}\rangle, \ldots |{M_{i,d}}}\rangle \right\}$) by the entries of $M_i$ (or $M_{i+1}$) with a given $H$ \cite{Zhang2020}. Then, with the known entries of $M_1$ and $M_{2k}$ and a given $H$ , we can uniquely determine all other sets of measurements, $M_2, M_3,\dots, M_J$ and $ M_{J+1},M_{J+2},\dots, M_{2k-1}$ from the ladder derivation of Eq. (3). Mathematically, this allows us to use numerical optimization to obtain the maximal value of ${P_1} - {P_2}$ for a given \(J\) by searching proper  matrices \(H\). Specifically, by utilizing the function \texttt{NMaximize} in Mathematica with the parameter \texttt{WorkingPrecision} set to 20, we obtain the maximum Cabello's fraction, as shown in Table~\ref{tab:table1} and~\ref{tab:table2}, and the optimal states are given in Appendix C. We can see that the maximum Cabello's fraction increases with dimension $d$ and setting $k$. Furthermore, in comparison with Hardy's fraction, the Increasing fraction increases with dimension $d$ and decreases with settings $k$. It is noteworthy that \(J = 1\) is the optimal choice to achieve a larger Cabello's fraction (see Appendix B for more details). For example, the maximum Cabello's fraction in the (2, 2, 1) scenario is calculated as 0.1250, surpassing the value obtained in the original Cabello's version \cite{Kunkri2006} in the (2, 2, 2) scenario, which is 0.1078. Consequently, both our theoretical and experimental results are involve in \((k, d, 1)\) scenario. 

\begin{table}
\caption{\label{tab:table1}%
Theoretical Cabello's fraction, Hardy's fraction and their comparison for ($k$, 2, 1) scenario}
\begin{ruledtabular}
\begin{tabular}{lcccc}
 Setting&\textrm{$k=3$}&
\textrm{$k=4$}&
\textrm{$k=5$}&
\textrm{$k=6$}\\
\hline
 Cabello's fraction&0.207107& 0.259733& 0.295755& 0.321900\\
 Hardy's fraction& 0.174550& 0.231263& 0.270880&0.299953\\
 Increasing fraction& 0.032557& 0.028470& 0.024875&0.021947\\
\end{tabular}
\end{ruledtabular}
\end{table}
\begin{table}
\caption{\label{tab:table2}%
Theoretical Cabello's fraction, Hardy's fraction and their comparison for (2, $d$, 1) scenario}
\begin{ruledtabular}
\begin{tabular}{lccc}
 Dimension&\textrm{$d=2$}&
\textrm{$d=3$}&
\textrm{$d=4$}\\
\hline
 Cabello's fraction& 0.125000& 0.193093& 0.238389\\
 Hardy's fraction& 0.090170& 0.141327&0.176512\\
 Increasing fraction& 0.034830& 0.051766&0.061877\\
\end{tabular}
\end{ruledtabular}
\end{table}

\section{Experimental setup and results}
\begin{figure*}[t]
\includegraphics[width=0.75\textwidth]{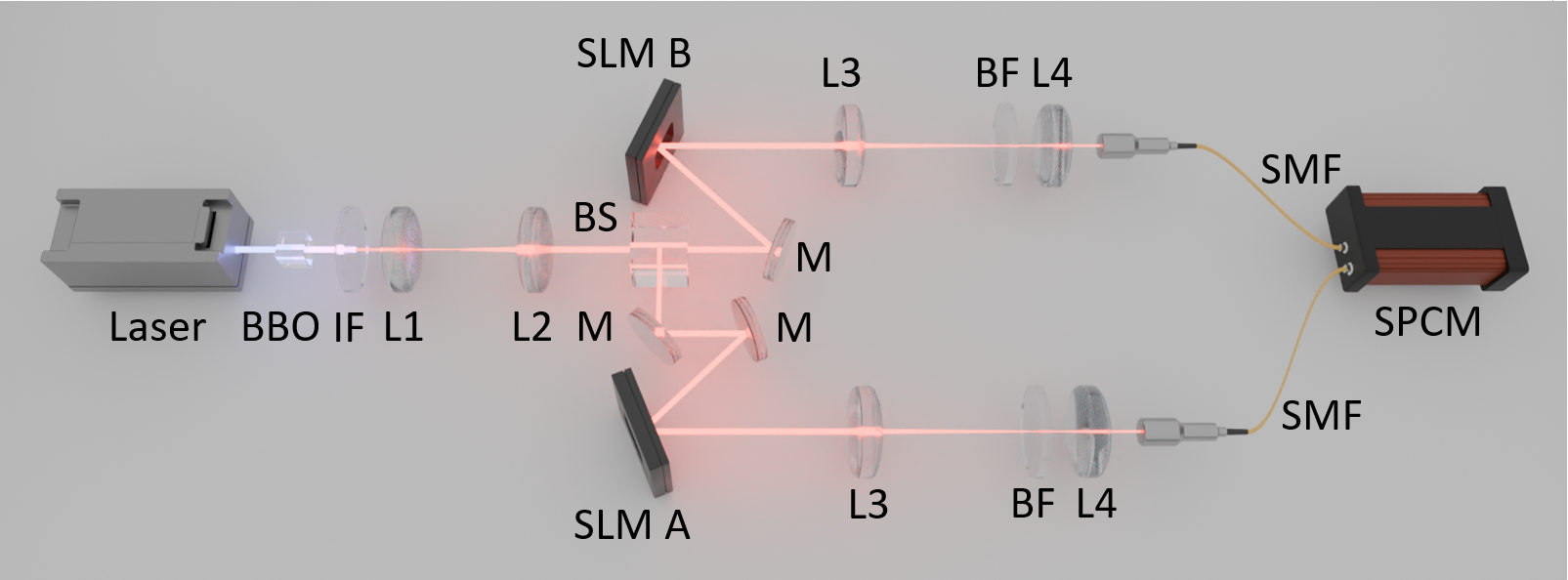}
\caption{\label{fig:setup} Experimental setup for demonstrating Cabello-type paradox with high-dimensional OAM entanglement. A mode-locked 355 nm ultraviolet laser serves as the pump for a 3-mm-thick $\beta$-barium borate (BBO) crystal, generating collinear frequency-degenerate photon pairs at 710 nm. A long pass filter (IF) behind the crystal blocks the pump beam, followed by a nonpolarizing beam splitter (BS) to separate the signal and idler photons. In each down-converted arm, a 4-f telescope consisting of two lenses (L1, f1=200mm and L2, f2=400mm) images the output facet of the BBO onto two SLMs (SLM A and SLM B, CAS MICROSTAR, FSLM-2K70-VIS). These SLMs are loaded with computer-designed holograms for the preparation of desired OAM measurement states and for implementing entanglement concentration procedures. Subsequently, another telescope (L3, f3=500mm and L4, f4=4mm) is used to reimage the plane of the SLM onto the input facet of a single-mode fiber (SMF) connected to a single photon counting module (Excelitas, SPCM-AQRH-14-FC) . In addition, two bandpass filters (BF) with a bandwidth of 10 nm and a center wavelength of 710 nm are placed in front of the SMF to reduce the detection of noise photons. The outputs of the two single-photon counters are connected to a coincidence counting circuit with a 25 ns coincidence time window.}
\end{figure*}
\begin{figure*}[t]
\includegraphics[width=0.75\textwidth]{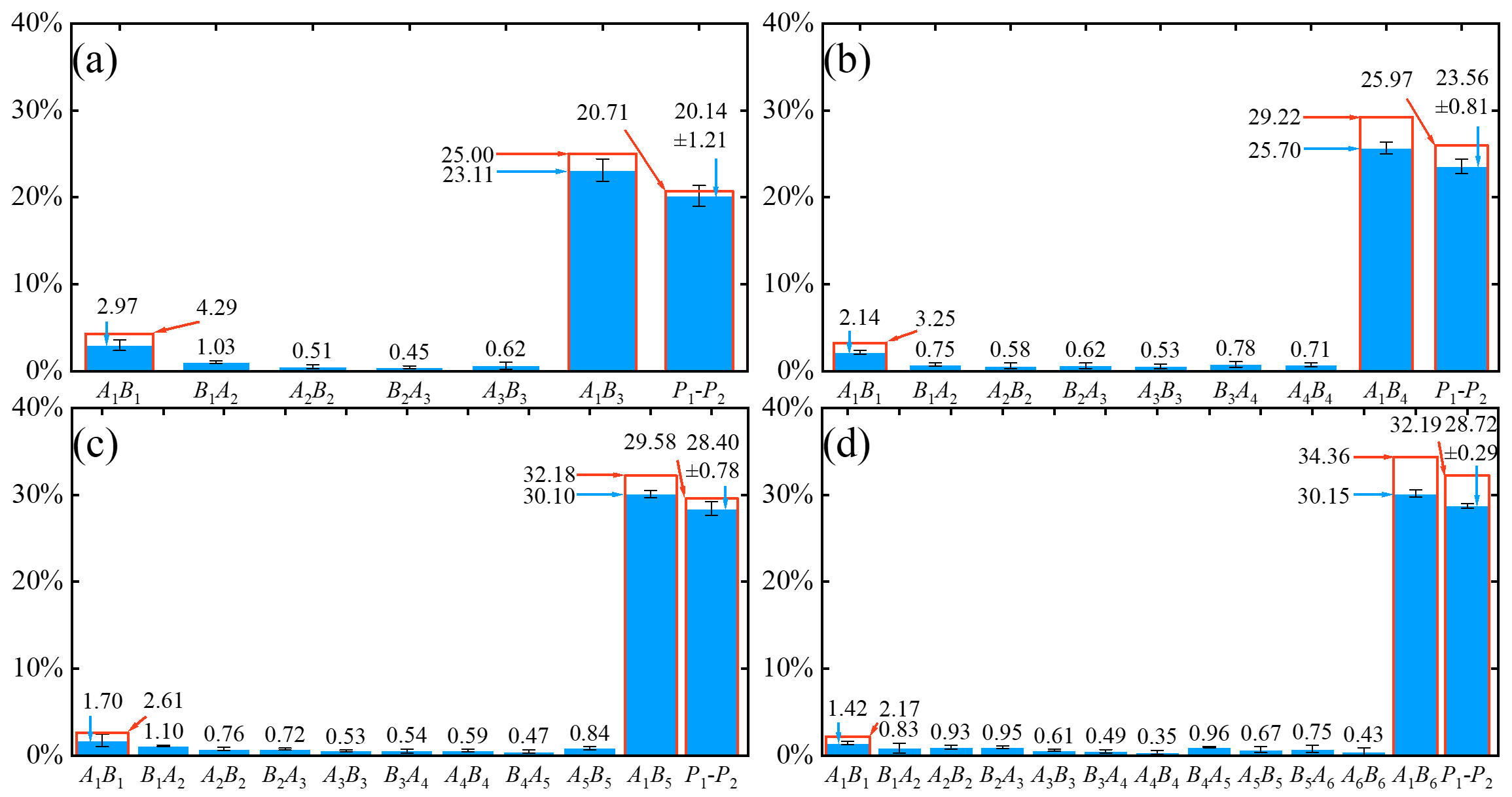}
\caption{\label{fig:result_1} Cabello-type paradox in ($k$, 2, 1) scenario: (a) $H^{\rm{opt}}_{(3,2,1)}$, (b) $H^{\rm{opt}}_{(4,2,1)}$, (c) $H^{\rm{opt}}_{(5,2,1)}$, and $H^{\rm{opt}}_{(6,2,1)}$. The empty bars (red edges) are the theoretical values while the solid bars (blue) are the experimental results. The error bars are the standard deviation of the experimental results. $A_iB_j$ and $B_jA_i$ stand for $P(A_i > B_j )$ and $P(B_j > A_i )$, respectively.}
\end{figure*}
To facilitate experimental verification, the optimal states given in Appendix C need to be transformed into more compact forms via Schmidt decomposition \cite{Mair2001,Nielsen2000}. According to the Schmidt decomposition theorem, any bipartite qudit state can be represented as $\left| \varphi  \right\rangle  = \sum\nolimits_g {{\lambda _g}{{\left| g \right\rangle }_A}{{\left| g \right\rangle }_B}}$. This mathematical procedure involves a singular value decomposition, i.e., $H = U\Sigma {V^{\dag}}$ , where \(\Sigma\) is 
a diagonal matrix with $ {\Sigma _{gg}}={\lambda _g} $ and unitary matrices \(U\) and \(V\) satisfy ${\left| g \right\rangle _A} = \sum_a {{U_{ag}}{{\left| a \right\rangle }_A}}$, ${\left| g \right\rangle _B} = \sum_b {V_{gb}^\dag {{\left| b \right\rangle }_B}}$ \cite{Nielsen2000}. Consequently, the measurement bases can be reconstructed as: \(|M_{i,s} \rangle = \sum_{g=1}^d m^\prime_{i,s,g} |g\rangle_A\) when $i$ is odd and \(|M_{i,s} \rangle = \sum_{g=1}^d m^\prime_{i,s,g} |g\rangle_B\) when $i$ is even. Here, $m^\prime_{i,s,g}$ can be obtained through straightforward algebraic operations. Employing this method, we derive the optimal quantum states as 
\begin{subequations}
\label{eq:7}
\begin{align}
H_{(2,2,1)}^{\text{opt}} = \rm{diag}&(0.866025,0.500000),\\
H_{(3,2,1)}^{\text{opt}} = \rm{diag}&(0.840896,0.541196),\\
H_{(4,2,1)}^{\text{opt}} = \rm{diag}&(0.821767,0.569823),\\
H_{(5,2,1)}^{\text{opt}} = \rm{diag}&(0.807542,0.589811),\\
H_{(6,2,1)}^{\text{opt}} = \rm{diag}&(0.796654,0.604435),\\
H_{(2,3,1)}^{\text{opt}} = \rm{diag}&(0.802376,0.456065,0.384963),\\
H_{(2,4,1)}^{\text{opt}} = \rm{diag}&(0.762167,0.432447,0.357871,0.322521),
\end{align}
\end{subequations}
with their corresponding measurement states detailed in Appendix D.

The experimental setup is shown in Fig.~\ref{fig:setup}. Taking advantage of the reconfigurable properties of the spatial light modulator (SLM), this configuration has been used to demonstrate Hardy’s paradox \cite{Zhang2020,Chen2017} and versions of the EPR paradox in both angular and radial domains \cite{Leach2010,Chen2019}. In spontaneous parametric down-conversion (SPDC), the two-photon OAM entangled state can be written as $\sum\nolimits_\ell {{{\rm{C}}_\ell}{{\left| \ell \right\rangle }_A}} {\left| { - \ell} \right\rangle _B}$, where $C_\ell$ denote the probability amplitude of signal photon with $\ell\hbar$ OAM and idler photon with $-\ell\hbar$ OAM \cite{Mair2001}. Note that the OAM eigenstates naturally form an orthogonal and complete basis, playing an ideal candidate to represent the Schmidt basis. Therefore, we can construct a larger but finite OAM subspace to formulate the Cabello-type paradox. In a special $d$-dimensional OAM subspace, the measurement bases can be redefined as $\left| {{M_{i,s}}} \right\rangle = \sum_{g = 1}^d {{m^\prime_{i,s,g}}{{\left| {{\ell_g}} \right\rangle }_A}} $ if $i$ is odd or $\left| {{M_{i,s}}} \right\rangle = \sum_{g = 1}^d {{m^\prime_{i,s,g}}{{\left| { - {\ell_g}} \right\rangle }_B}} $ if $i$ is even, and the quantum state can be redefined as $\left| \psi  \right\rangle  = \sum_{g = 1}^d {{\lambda _g}{{\left| {{\ell_g}} \right\rangle }_A}{{\left| { - {\ell_g}} \right\rangle }_B}}$, all of $\lambda _g$ as shown in Eq. (7).

To experimentally generate the optimal states, the OAM entangled state produced by SPDC needs to be tailored through the so-called Procrustean method of entanglement concentration \cite{Bennett1996}. That is, the weight amplitudes of the OAM modes with excessively high probability amplitudes in the initial quantum state need to be reduced to be consistent with the amplitudes of the optimal quantum state. This is achieved by changing the diffraction efficiency of blazed phase grating on the SLM. By reducing the contrast of blazed phase grating, lower diffraction efficiency for the OAM modes can be obtained, reducing their weight amplitudes. Through this method, we experimentally prepare the optimal states in Eq.~(\ref{eq:7}). For instance, in the (2,3,1) scenario, we focus on the three-dimensional OAM subspace spanned by $\ell_1,\ell_2,\ell_3 = 0,+1,-1$. The quantum state product via SPDC can be expressed as $\left| {\psi _3^{{\text{SPDC}}}} \right\rangle  = 0.848611{\left| 0 \right\rangle _A}{\left| 0 \right\rangle _B} + 0.365378{\left| { + 1} \right\rangle _A}{\left| { - 1} \right\rangle _B} + 0.382569{\left| { - 1} \right\rangle _A}{\left| { + 1} \right\rangle _B}$. The optimal state for achieving the maximum Cabello's fraction is $\left| {\psi _{\left( {2,3,1} \right)}^{{\text{opt}}}} \right\rangle  = 0.802376{\left| 0 \right\rangle _A}{\left| 0 \right\rangle _B} + 0.456065{\left| { + 1} \right\rangle _A}{\left| { - 1} \right\rangle _B} + 0.384963{\left| { - 1} \right\rangle _A}{\left| { + 1} \right\rangle _B}$. To achieve this optimal state, we decrease the diffraction efficiencies of the ${\left| 0 \right\rangle _A}$ and ${\left| -1 \right\rangle _A}$ modes in SLM A, as well as the ${\left| 0 \right\rangle _B}$ and ${\left| +1 \right\rangle _B}$ modes on SLM B. Specifically, we set the diffraction efficiency to ${\eta _0}$ for both ${\left| 0 \right\rangle _A}$ and ${\left| 0 \right\rangle _B}$ modes, and ${\eta _1}$ for both ${\left| -1 \right\rangle _A}$ and ${\left| +1 \right\rangle _B}$ modes. For each combination of ${\eta _0}$ and ${\eta _1}$, we sequentially load the modes ${\left| 0 \right\rangle _A}{\left| 0 \right\rangle _B}$, ${\left| +1 \right\rangle _A}{\left| -1 \right\rangle _B}$, and ${\left| -1 \right\rangle _A}{\left| +1 \right\rangle _B}$ onto SLMs and record the coincidence counts. After multiple adjustments, we successfully prepare the optimal state in terms of the data of coincidence counts. For measurements, we first calculate the desired measurement bases by multiplying them (see Appendix D) with the indices of the diffraction coefficients. Next, we sequentially load the OAM superposition states corresponding to these modified bases onto SLMs and record the coincidence counts, with which we obtain the experimental results of the CNA.

\begin{figure}[b]
\includegraphics[width=0.45\textwidth]{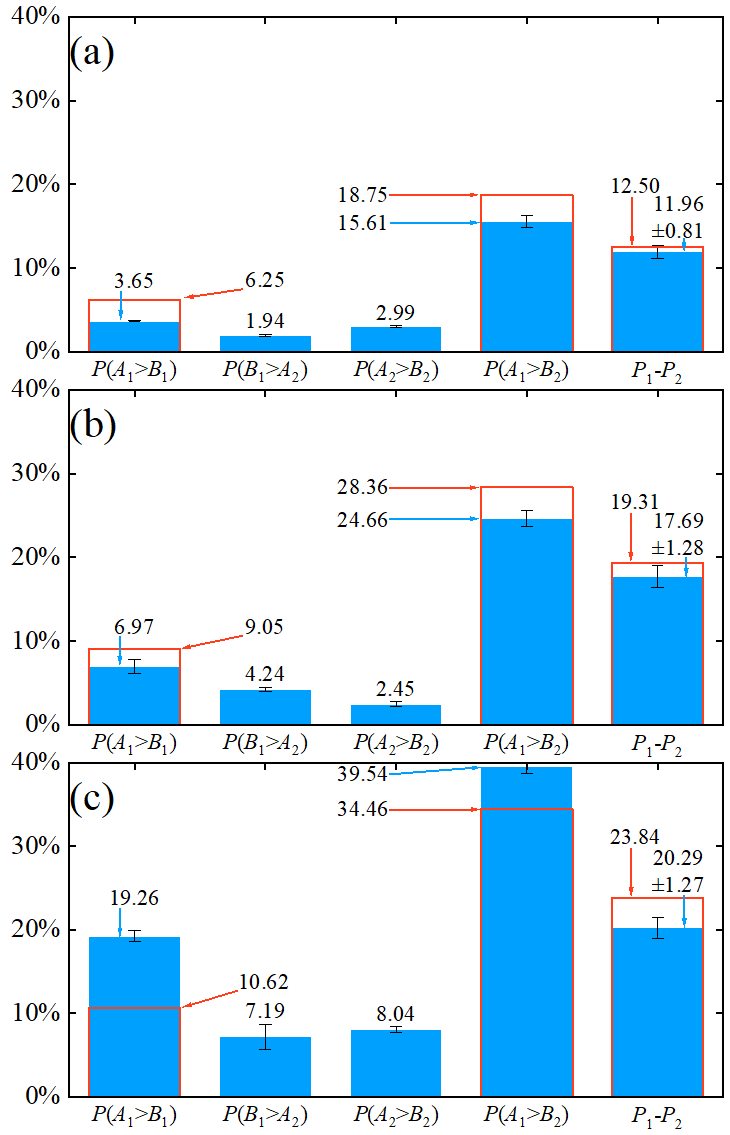}
\caption{\label{fig:result_2} Cabello-type paradox in (2, $d$, 1) scenario: (a) $H^{\rm{opt}}_{(2,2,1)}$, (b) $H^{\rm{opt}}_{(2,3,1)}$, and (c) $H^{\rm{opt}}_{(2,4,1)}$. The empty bars (red edges) are the theoretical values while the solid bars (blue) are the experimental results.}
\end{figure}

In the first set of experiments, our focus lies within the two-dimensional OAM subspace spanned by $\ell_1,\ell_2=+1,-1$, corresponding to ($k$, 2, 1) scenarios with measurement settings $k$ ranging from 3 to 6. For clarity, we denote ${M_{2i - 1}}$ as ${A_i}$ (Alice), and ${M_{2i}}$ as ${B_i}$ (Bob). Then CNA can be translated as:
\begin{equation}
\label{eq:8}
    P\left( {{A_1} > {B_k}} \right) = {P_1},
\end{equation}
\begin{equation}
\label{eq:9}
    P\left( {{A_1} > {B_1}} \right) = {P_2},
\end{equation}
\begin{equation}
\label{eq:10}
    P\left( {{B_i} > {A_{i+1}}} \right) = 0,{\rm{ for }}\ i = 1, \ldots, k-1,
\end{equation}
\begin{equation}
\label{eq:11}
    P\left( {{A_{i+1}} > {B_{i+1}}} \right) = 0,{\rm{ for }}\ i = 1, \ldots, k-1.
\end{equation}
The experimental results are illustrated in Fig.~\ref{fig:result_1}. We can see that the observed Cabello's fractions can reach 20.14±1.21\%, 23.56±0.81\%, 28.40±0.78\% and 28.72±0.29\%, respectively. One can see that these results show a good agreement with the predictions in Table~\ref{tab:table1}. Moreover, the probabilities $P\left( {{B_i} > {A_{i+1}}} \right)$ and $P\left( {{A_{i+1}} > {B_{i+1}}} \right)$ with $i$ range from 1 to $k-1$ are nearly close to zero, align with theoretical predictions in Eqs.~(\ref{eq:10})-(\ref{eq:11}). As the local realism dictates that in such cases Cabello's fractions ${P_1} - {P_2} \leq 0$, our experimental results demonstrate a violation of local realism by quantum mechanics. Remarkably, these results significantly surpass the bound that was limited in original version \cite{Rai2021}.

In the second set of experiments, we consider (2, $d$, 1) scenarios with the dimension $d$ being 2, 3, 4. Their corresponding OAM subspaces are spanned by \(\ell_1,\ell_2 = +1,-1\); \(\ell_1,\ell_2,\ell_3 = 0,+1,-1\); and \(\ell_1,\ell_2,\ell_3,\ell_4 = 0,+1,-1,+2\); respectively. The experimental results are shown in Fig.~\ref{fig:result_2}. Notably, the successful probabilities are 11.96±0.81\%, 17.69±1.28\%, and 20.29±1.27\%, respectively, demonstrating a trend of increase with dimension and a reasonable agreement with the quantum-mechanical predictions. All of the experimental observations show higher successful probabilities than those of Hardy's paradox \cite{Zhang2020}.

\section{Discussion and conclusion}
Similar to the Bell experiments, testing the CNA is also affected by the detection \cite{vertesi2010closing} and locality loopholes \cite{hensen2015loophole}, with CNA requiring a higher detection efficiency \cite{li2018test,zhao2024loophole}. Notably, in our experiment, neither detection nor locality loopholes were closed, given our overall experimental efficiency less than 9\% and the switching time of our measurement device (SLM) with a response time of tens of milliseconds. Recent studies suggest that high-dimensional systems, e.g., our extended high-dimensional multi-setting CNA theory, can significantly mitigate the detection efficiency requirements for loophole-free experiments \cite{hu2022high}. However, achieving genuinely loophole-free tests of locality is not the direct aim of our experiment. Instead, we employ high-dimensional OAM entanglement to verify the new version of CNA. Additionally, decoherence-induced losses during photon propagation can also cause deviations of the experimental results from the theoretical values \cite{sabuncu2007nonunity,lassen2010experimental}. Consequently, these imperfections in the measurement process lead to non-zero probabilities for undesired outcomes in our experimental results. It is natural to resort to the Clauser-Horne inequality. Assuming the events ${M_{2k}} < {M_1}$ and ${M_i} \le {M_{i + 1}},{\rm{for }} \ i = 1,2, \ldots ,2k-1$ occur, from \textit{Theorem 1}, we can deduce that the event ${M_{2k}} < {M_1} \cap \left( { \cap _{i = 1}^{2k-1}{M_i} \le {M_{i + 1}}} \right)$ is impossible. According to Ref. \cite{Abramsky2012}, certain events can be translated into a logical Bell inequality if their intersection is not possible. This inequality can be expressed as $\sum\nolimits_i {{p_i}} \le N - 1$, where $p_i$ represents the probability of each event and $N$ represents the number of events. Thus, in our context, we can translate the aforementioned impossible event as $P\left( {{M_{2k}} < {M_1}} \right) + \sum\nolimits_{i = 1}^{2k - 1} {P\left( {{M_i} \le {M_{i + 1}}} \right)} \le 2k - 1.$ It is noteworthy that this inequality is equivalent to a Clauser-Horne inequality \cite{Mermin1994} encompassed by the general Hardy’s inequality \cite{Meng2018}, and also equivalent to a n-cycle noncontextuality inequality $\sum\nolimits_{i = 0}^{n - 1} {\left( {1 - 2{\delta _{i0}}} \right)\left\langle {{M_i}{M_{i + 1}}} \right\rangle } \le n - 2$ when $n=2k$ and $d=2$ if $P\left( {{M_i} > {M_{i + 1}}} \right) \equiv P\left( {{M_i} < {M_{i + 1}}} \right)$ (In the latter inequality, the eigenvalues of observables is ±1) \cite{Araujo2013}. Now, we denote the Bell expression $S$ as 
\begin{equation}
S = P\left( {{M_{2k}} < {M_1}} \right) + \sum\limits_{i = 1}^{2k - 1} {P\left( {{M_i} \le {M_{i + 1}}} \right)} -(2k - 1)\le0.
\end{equation}
Then, based on the experimental results in Fig. 3 and Fig. 4, the experimental $S$ can be calculated as: $S^{\rm{opt}}_{(3,2,1)}=17.54\pm1.22\%$, $S^{\rm{opt}}_{(4,2,1)}=19.59\pm1.52\%$, $S^{\rm{opt}}_{(5,2,1)}=22.85\pm1.08\%$, $S^{\rm{opt}}_{(6,2,1)}=21.76\pm0.85\%$, $S^{\rm{opt}}_{(2,2,1)}=7.02\pm1.07\%$, $S^{\rm{opt}}_{(2,3,1)}=10.99\pm1.18\%$, and $S^{\rm{opt}}_{(2,4,1)}=5.06\pm2.09\%$. We can see the experimental results can violate the inequality up to 26 standard deviations, and thus confirm the nonlocal behavior of quantum mechanics.

In summary, we have successfully extended CNA to arbitrary ($k$, $d$) scenarios, combining theoretical insights with experimental validation. The theoretical analysis, supported by compatibility graphs, establishes a novel logical framework for CNA, demonstrating an increased maximum successful probability correlated with settings ($k$) and dimensions ($d$). In alignment with our theoretical exploration, our experimental efforts utilize controllable photonic orbital angular momentum entanglement, revealing nonlocality with experimentally recorded probabilities of 20.29\% and 28.72\% in the (2, 4, 1) and (6, 2, 1) scenarios, respectively. These results underscore an intensified contradiction between quantum mechanics and classical theory, highlighting the significance of CNA in high-dimensional scenarios. Intriguingly, our analysis establishes that, even in high-dimensional systems, the CNA can reach a larger successful probability than the general Hardy’s paradox. Since higher success probability implies greater robustness and noise resistance, our scheme is better equipped to withstand the effects of environmental noise in high-dimensional quantum communication. Recently, we became aware that the paradox interpreted as the failure of the transitivity implications (FTI) and the Cabello-Liang-Li (CLL) Hardy-type paradox, were also demonstrated theoretically \cite{chen2024hardy}. These are equivalent to the CNA in the $(k, d, 1)$ scenario and the $(k, d, k)$ scenario, respectively. Their theoretical results for the FTI-based Hardy-type paradox, as well as their earlier work \cite{chen2023quantum}, align closely with those presented in Table I. In this regards, our results using high-dimensional OAM entanglement also offer a possible platform, from the experimental point of view, to demonstrate their theoretical findings. Furthermore, our findings contribute to the current understanding of quantum nonlocality, emphasizing the potential of CNA in device-independent applications \cite{zhang2022device,i2022quantum}, particularly for quantum systems with high-dimensional entanglement.

\begin{acknowledgments}
This work is supported by the National Natural Science Foundation of China (Grants No. 12034016, No. 12205107), the Natural Science Foundation of Fujian Province of China (Grant No. 2021J02002), the Program for New Century Excellent Talents in University of China (Grant No. NCET-13-0495), and the Natural Science Foundation of Xiamen Municipality (Grant No. 3502Z20227033).
\end{acknowledgments}

\appendix

\section{CABELLO-TYPE PARADOX DEGREE TO THE ORIGINAL CNA}

In the case of $(2, 2, 2)$, Cabello-type paradox degenerate to:
\begin{equation}
\label{eq:A1}
\begin{array}{l}
P\left( {{M_1} = 2,{M_4} = 1} \right) = {P_1},\\
P\left( {{M_1} = 2,{M_2} = 1} \right) = 0,\\
P\left( {{M_3} = 2,{M_4} = 1} \right) = 0,\\
P\left( {{M_2} = 2,{M_3} = 1} \right) = {P_2},\\
{P_1} - {P_2} > 0.
\end{array}
\end{equation}
Let observables $F=3-2M_3$, $G=2M_2-3$, $D=2M_1-3$, $E=3-2M_4$, then Eq.~(\ref{eq:A1}) is equivalent to:
\begin{equation}
    \begin{array}{l}
P\left( {F =  + 1,G =  + 1} \right) = {P_2},\\
P\left( {D =  + 1,G =  - 1} \right) = 0,\\
P\left( {F =  - 1,E =  + 1} \right) = 0,\\
P\left( {D =  + 1,E =  + 1} \right) = {P_1},\\
{P_1} - {P_2} > 0.
\end{array}
\end{equation}
It’s CNA for two particles \cite{Kunkri2006}.

\section{THE PROOF OF THE MAXIMUM CABELLO’S FRACTION IS ACHIEVED WHEN J = 1}
To streamline our theory and its computation, we present the subsequent theorem by interchanging the order and outcomes of measurements, akin to reference \cite{Meng2018}:

\textit{Theorem 2}. Cabello-type paradox for general ($k$, $d$, $J$) scenario is equivalent to ($k$, $d$, $2k-J$) scenario.

\textit{Proof}. Put observables $M_i = \overline{M'_{2k+1-i}}$ for arbitrary $i$, where the outcome of $\overline{M'_{2k+1-i}}$ is equal to $m$ if the outcome of $M'_{2k+1-i}$ is $d-m$. It means $P(M_i=m, M_{i'}=m') = P(\overline{M'_{2k+1-i}} = d-m, \overline{M'_{2k+1-i'}} = d-m')$, so we have
\begin{equation}
\begin{array}{l}
P\left( {{M_i} > {M_{i'}}} \right)\\
 = \sum\limits_{m > m'} {P\left( {{M_i} = m,{M_{i'}} = m'} \right)} \\
 = \sum\limits_{d - m < d - m'} {P\left( {{{M'}_{2k + 1 - i}} = d - m,{{M'}_{2k + 1 - i'}} = d - m'} \right)} \\
 = P\left( {{{M'}_{2k + 1 - i}} < {{M'}_{2k + 1 - i'}}} \right),
\end{array}
\end{equation}
then Eqs.~(\ref{eq:1})-(\ref{eq:4}) has the alternative form:
\begin{equation}
\begin{array}{l}
P\left( {{{M'}_{2k}} < {{M'}_1}} \right) = {P_1},\\
P\left( {{{M'}_{2k + 1 - J}} < {{M'}_{2k - J}}} \right) = {P_2},\\
P\left( {{{M'}_{2k + 1 - i}} < {{M'}_{2k - i}}} \right) = 0,{\rm{ for }}\ i = 1,2, \ldots 2k-1,\ i \ne J,\\
{P_1} - {P_2} > 0,
\end{array}
\end{equation}
which is the Cabello-type paradox for ($k$, $d$, $2k-J$) scenario.

According to \textit{Theorem 2}, our focus is narrowed down to scenarios where $J \leq k$. By using aforementioned numerical strategy, we discover that setting $J=1$ yields the maximum Cabello's fraction. For instance, in the case of $(5, 2, J)$ scenario, the corresponding maximum Cabello's fractions for $J=1,2,\ldots,5$ are sequentially 0.295755, 0.284323, 0.278595, 0.276103, and 0.275415.

\section{THE OPTIMAL QUANTUM STATES}

The optimal quantum states directly computed by optimization algorithm are:
\[\begin{array}{l}
H_{(2,2,1)}^{\rm{opt}} = \left( \begin{array}{cc}
0.750000 & 0.433013 \\
-0.250000 & 0.433013
\end{array} \right),\\[10pt]
H_{(3,2,1)}^{\rm{opt}} = \left( \begin{array}{cc}
-0.776887 & 0.321797 \\
-0.207107 & -0.500000
\end{array} \right),\\[10pt]
H_{(4,2,1)}^{\rm{opt}} = \left( \begin{array}{cc}
-0.779580 & 0.259917 \\
-0.180229 & -0.540570
\end{array} \right),\\[10pt]
H_{(5,2,1)}^{\rm{opt}} = \left( \begin{array}{cc}
-0.776705 & 0.221028 \\
-0.161434 & -0.567288
\end{array} \right),\\[10pt]
H_{(6,2,1)}^{\rm{opt}} = \left( \begin{array}{cc}
-0.772609 & 0.194250 \\
-0.147381 & -0.586192
\end{array} \right),\\[10pt]
H_{(2,3,1)}^{\rm{opt}} = \left( \begin{array}{ccc}
-0.633325 & 0.359393 & 0.309398 \\
-0.173969 & -0.372389 & -0.134408 \\
-0.186571 & 0.159324 & -0.356106
\end{array} \right),\\[10pt]
H_{(2,4,1)}^{\rm{opt}} = \left( \begin{array}{cccc}
0.562093 & 0.316214 & -0.266562 & 0.249385 \\
-0.137751 & 0.328611 & -0.131867 & 0.080517 \\
-0.132708 & -0.123247 & -0.322726 & 0.110432 \\
0.156836 & 0.118696 & -0.125494 & -0.310479
\end{array} \right).
\end{array}\]

\section{THE DESIRED OAM MEASUREMENT STATES}

In the first experiment, we choose the OAM modes \(\ell_1 = +1, \ell_2 = -1\) for $(k,2,2)$ scenario. The desired OAM measurement states in the (3, 2, 1) scenario are
\begin{align*}
|A_{1,1}\rangle &= \begin{bmatrix}-1 \\ 0\end{bmatrix}, &
|A_{1,2}\rangle &= \begin{bmatrix}0 \\ 1\end{bmatrix}, \\
|A_{2,1}\rangle &= \begin{bmatrix}-0.840896 \\ 0.541196\end{bmatrix}, &
|A_{2,2}\rangle &= \begin{bmatrix}-0.541196 \\ -0.840896\end{bmatrix}, \\
|A_{3,1}\rangle &= \begin{bmatrix}-0.541196 \\ 0.840896\end{bmatrix}, &
|A_{3,2}\rangle &= \begin{bmatrix}-0.840896 \\ -0.541196\end{bmatrix}, \\
|B_{1,1}\rangle &= \begin{bmatrix}0.923880 \\ -0.382683\end{bmatrix}, &
|B_{1,2}\rangle &= \begin{bmatrix}-0.382683 \\ -0.923880\end{bmatrix}, \\
|B_{2,1}\rangle &= \begin{bmatrix}-0.707107 \\ 0.707107\end{bmatrix}, &
|B_{2,2}\rangle &= \begin{bmatrix}-0.707107 \\ -0.707107\end{bmatrix}, \\
|B_{3,1}\rangle &= \begin{bmatrix}-0.382683 \\ 0.923880\end{bmatrix}, &
|B_{3,2}\rangle &= \begin{bmatrix}-0.923880 \\ -0.382683\end{bmatrix}.
\end{align*}
The desired OAM measurement states in the (4, 2, 1) scenario are
\begin{align*}
|A_{1,1}\rangle &= \begin{bmatrix}-1 \\ 0\end{bmatrix}, &|A_{1,2}\rangle &= \begin{bmatrix}0 \\ 1\end{bmatrix}, \\
|A_{2,1}\rangle &= \begin{bmatrix}-0.901235 \\ 0.433331\end{bmatrix}, &|A_{2,2}\rangle &= \begin{bmatrix}-0.433331 \\ -0.901235\end{bmatrix}, \\
|A_{3,1}\rangle &= \begin{bmatrix}-0.707107 \\ 0.707107\end{bmatrix}, &|A_{3,2}\rangle &= \begin{bmatrix}-0.707107 \\ -0.707107\end{bmatrix}, \\
|A_{4,1}\rangle &= \begin{bmatrix}-0.433331 \\ 0.901235\end{bmatrix}, &|A_{4,2}\rangle &= \begin{bmatrix}-0.901235 \\ -0.433331\end{bmatrix}, \\
|B_{1,1}\rangle &= \begin{bmatrix}0.948663 \\ -0.316290\end{bmatrix}, &|B_{1,2}\rangle &= \begin{bmatrix}-0.316290 \\ -0.948663\end{bmatrix}, \\
|B_{2,1}\rangle &= \begin{bmatrix}-0.821767 \\ 0.569823\end{bmatrix}, &|B_{2,2}\rangle &= \begin{bmatrix}-0.569823 \\ -0.821767\end{bmatrix}, \\
|B_{3,1}\rangle &= \begin{bmatrix}-0.569823 \\ 0.821767\end{bmatrix}, &|B_{3,2}\rangle &= \begin{bmatrix}-0.821767 \\ -0.569823\end{bmatrix}, \\
|B_{4,1}\rangle &= \begin{bmatrix}-0.316290 \\ 0.948663\end{bmatrix}, &|B_{4,2}\rangle &= \begin{bmatrix}-0.948663 \\ -0.316290\end{bmatrix}.
\end{align*}
The desired OAM measurement states in the (5, 2, 1) scenario are
\begin{align*}
|A_{1,1}\rangle &= \begin{bmatrix}-1 \\ 0\end{bmatrix}, &
|A_{1,2}\rangle &= \begin{bmatrix}0 \\ 1\end{bmatrix}, \\
|A_{2,1}\rangle &= \begin{bmatrix}-0.931774 \\ 0.363039\end{bmatrix}, &
|A_{2,2}\rangle &= \begin{bmatrix}-0.363039 \\ -0.931774\end{bmatrix}, \\
|A_{3,1}\rangle &= \begin{bmatrix}-0.807542 \\ 0.589811\end{bmatrix}, &
|A_{3,2}\rangle &= \begin{bmatrix}-0.589811 \\ -0.807542\end{bmatrix}, \\
|A_{4,1}\rangle &= \begin{bmatrix}-0.589811 \\ 0.807542\end{bmatrix}, &
|A_{4,2}\rangle &= \begin{bmatrix}-0.807542 \\ -0.589811\end{bmatrix}, \\
|A_{5,1}\rangle &= \begin{bmatrix}-0.363039 \\ 0.931774\end{bmatrix}, &
|A_{5,2}\rangle &= \begin{bmatrix}-0.931774 \\ -0.363039\end{bmatrix}, \\
|B_{1,1}\rangle &= \begin{bmatrix}0.961814 \\ -0.273705\end{bmatrix}, &
|B_{1,2}\rangle &= \begin{bmatrix}-0.273705 \\ -0.961814\end{bmatrix}, \\
|B_{2,1}\rangle &= \begin{bmatrix}-0.882309 \\ 0.470670\end{bmatrix}, &
|B_{2,2}\rangle &= \begin{bmatrix}-0.470670 \\ -0.882309\end{bmatrix}, \\
|B_{3,1}\rangle &= \begin{bmatrix}-0.707107 \\ 0.707107\end{bmatrix}, &
|B_{3,2}\rangle &= \begin{bmatrix}-0.707107 \\ -0.707107\end{bmatrix}, \\
|B_{4,1}\rangle &= \begin{bmatrix}-0.470670 \\ 0.882309\end{bmatrix}, &
|B_{4,2}\rangle &= \begin{bmatrix}-0.882309 \\ -0.470670\end{bmatrix}, \\
|B_{5,1}\rangle &= \begin{bmatrix}-0.273705 \\ 0.961814\end{bmatrix}, &
|B_{5,2}\rangle &= \begin{bmatrix}-0.961814 \\ -0.273705\end{bmatrix}.
\end{align*}
The desired OAM measurement states in the (6, 2, 1) scenario are
\begin{align*}
|A_{1,1}\rangle &= \begin{bmatrix}-1 \\ 0\end{bmatrix}, &|A_{1,2}\rangle &= \begin{bmatrix}0 \\ 1\end{bmatrix}, \\
|A_{2,1}\rangle &= \begin{bmatrix}-0.949239 \\ 0.314555\end{bmatrix}, &|A_{2,2}\rangle &= \begin{bmatrix}-0.314555 \\ -0.949239\end{bmatrix}, \\
|A_{3,1}\rangle &= \begin{bmatrix}-0.866662 \\ 0.498896\end{bmatrix}, &|A_{3,2}\rangle &= \begin{bmatrix}-0.498896 \\ -0.866662\end{bmatrix}, \\
|A_{4,1}\rangle &= \begin{bmatrix}-0.707107 \\ 0.707107\end{bmatrix}, &|A_{4,2}\rangle &= \begin{bmatrix}-0.707107 \\ -0.707107\end{bmatrix}, \\
|A_{5,1}\rangle &= \begin{bmatrix}-0.498896 \\ 0.866662\end{bmatrix}, &|A_{5,2}\rangle &= \begin{bmatrix}-0.866662 \\ -0.498896\end{bmatrix}, \\
|A_{6,1}\rangle &= \begin{bmatrix}-0.314555 \\ 0.949239\end{bmatrix}, &|A_{6,2}\rangle &= \begin{bmatrix}-0.949239 \\ -0.314555\end{bmatrix}, \\
|B_{1,1}\rangle &= \begin{bmatrix}0.969818 \\ -0.243832\end{bmatrix}, &|B_{1,2}\rangle &= \begin{bmatrix}-0.243832 \\ -0.969818\end{bmatrix}, \\
|B_{2,1}\rangle &= \begin{bmatrix}-0.916407 \\ 0.400248\end{bmatrix}, &|B_{2,2}\rangle &= \begin{bmatrix}-0.400248 \\ -0.916407\end{bmatrix}, \\
|B_{3,1}\rangle &= \begin{bmatrix}-0.796654 \\ 0.604435\end{bmatrix}, &|B_{3,2}\rangle &= \begin{bmatrix}-0.604435 \\ -0.796654\end{bmatrix}, \\
|B_{4,1}\rangle &= \begin{bmatrix}-0.604435 \\ 0.796654\end{bmatrix}, &|B_{4,2}\rangle &= \begin{bmatrix}-0.796654 \\ -0.604435\end{bmatrix}, \\
|B_{5,1}\rangle &= \begin{bmatrix}-0.400248 \\ 0.916407\end{bmatrix}, &|B_{5,2}\rangle &= \begin{bmatrix}-0.916407 \\ -0.400248\end{bmatrix}, \\
|B_{6,1}\rangle &= \begin{bmatrix}-0.243832 \\ 0.969818\end{bmatrix}, &|B_{6,2}\rangle &= \begin{bmatrix}-0.969818 \\ -0.243832\end{bmatrix}.
\end{align*}

In the second experiment, we choose the OAM modes \(\ell_1 = +1, \ell_2 = -1\) for two-dimensional optimal state \(H^\text{opt}_{(2,2,1)}\); \(\ell_1 = 0, \ell_2 = +1, \ell_3 = -1\) for three-dimensional optimal state \(H^\text{opt}_{(2,3,1)}\); \(\ell_1 = 0, \ell_2 = +1, \ell_3 = -1, \ell_4 = +2\) for four-dimensional optimal state \(H^\text{opt}_{(2,4,1)}\). The desired OAM measurement states in the $(2, 2, 1)$ scenario are
\begin{align*}
|A_{1,1}\rangle &= \begin{bmatrix}1 \\ 0\end{bmatrix}, &
|A_{1,2}\rangle &= \begin{bmatrix}0 \\ 1\end{bmatrix}, \\
|A_{2,1}\rangle &= \begin{bmatrix}-0.707107 \\ 0.707107\end{bmatrix}, &
|A_{2,2}\rangle &= \begin{bmatrix}-0.707107 \\ -0.707107\end{bmatrix}, \\
|B_{1,1}\rangle &= \begin{bmatrix}0.866025 \\ -0.500000\end{bmatrix}, &
|B_{1,2}\rangle &= \begin{bmatrix}0.500000 \\ 0.866025\end{bmatrix}, \\
|B_{2,1}\rangle &= \begin{bmatrix}-0.500000 \\ 0.866025\end{bmatrix}, &
|B_{2,2}\rangle &= \begin{bmatrix}-0.866025 \\ -0.500000\end{bmatrix}.
\end{align*}
The desired OAM measurement states in the (2, 3, 1) scenario are
\begin{align*}
|A_{1,1}\rangle &= \begin{bmatrix}-0.981861 \\ 0 \\ -0.189603\end{bmatrix}, &
|A_{1,2}\rangle &= \begin{bmatrix}0.134070 \\ 0.707107 \\ -0.694280\end{bmatrix}, \\
|A_{1,3}\rangle &= \begin{bmatrix}-0.134070 \\ 0.707107 \\ 0.694280\end{bmatrix}, &
|A_{2,1}\rangle &= \begin{bmatrix}0.558726 \\ -0.707107 \\ 0.433389\end{bmatrix}, \\
|A_{2,2}\rangle &= \begin{bmatrix}0.612904 \\ 0 \\ -0.790157\end{bmatrix}, &
|A_{2,3}\rangle &= \begin{bmatrix}0.558726 \\ 0.707107 \\ 0.433389\end{bmatrix}, \\
|B_{1,1}\rangle &= \begin{bmatrix}0.777099 \\ -0.559000 \\ 0.289199\end{bmatrix}, &
|B_{1,2}\rangle &= \begin{bmatrix}-0.528631 \\ -0.330348 \\ 0.781933\end{bmatrix}, \\
|B_{1,3}\rangle &= \begin{bmatrix}-0.341564 \\ -0.760519 \\ -0.552217\end{bmatrix}, &
|B_{2,1}\rangle &= \begin{bmatrix}0.341564 \\ -0.760519 \\ 0.552217\end{bmatrix}, \\
|B_{2,2}\rangle &= \begin{bmatrix}0.528631 \\ -0.330348 \\ -0.781933\end{bmatrix}, &
|B_{2,3}\rangle &= \begin{bmatrix}-0.777099 \\ -0.559000 \\ -0.289199\end{bmatrix}.
\end{align*}
The desired OAM measurement states in the (2, 4, 1) scenario are
\begin{align*}
|A_{1,1}\rangle &= \begin{bmatrix}0.964350 \\ 0 \\ 0.264629 \\ 0\end{bmatrix}, &
|A_{1,2}\rangle &= \begin{bmatrix}0.187121 \\ 0.477609 \\ -0.681899 \\ 0.521431\end{bmatrix}, \\
|A_{1,3}\rangle &= \begin{bmatrix}0 \\ 0.737414 \\ 0 \\ -0.675441\end{bmatrix}, &
|A_{1,4}\rangle &= \begin{bmatrix}0.187121 \\ -0.477609 \\ -0.681899 \\ -0.521431\end{bmatrix}, \\
|A_{2,1}\rangle &= \begin{bmatrix}0.469000 \\ -0.636951 \\ 0.529187 \\ -0.307072\end{bmatrix}, &
|A_{2,2}\rangle &= \begin{bmatrix}-0.529187 \\ 0.307072 \\ 0.469000 \\ -0.636951\end{bmatrix}, \\
|A_{2,3}\rangle &= \begin{bmatrix}0.529187 \\ 0.307072 \\ -0.469000 \\ -0.636951\end{bmatrix}, &
|A_{2,4}\rangle &= \begin{bmatrix}0.469000 \\ 0.636951 \\ 0.529187 \\ 0.307072\end{bmatrix}, \\
|B_{1,1}\rangle &= \begin{bmatrix}0.715887 \\ -0.551646 \\ 0.379278 \\ -0.198344\end{bmatrix}, &
|B_{1,2}\rangle &= \begin{bmatrix}0.509916 \\ 0.021673 \\ -0.618488 \\ 0.597485\end{bmatrix}, \\
|B_{1,3}\rangle &= \begin{bmatrix}-0.400459 \\ -0.557354 \\ 0.293274 \\ 0.665567\end{bmatrix}, &
|B_{1,4}\rangle &= \begin{bmatrix}0.259083 \\ 0.620139 \\ 0.622585 \\ 0.400864\end{bmatrix}, \\
|B_{2,1}\rangle &= \begin{bmatrix}0.259083 \\ -0.620139 \\ 0.622585 \\ -0.400864\end{bmatrix}, &
|B_{2,2}\rangle &= \begin{bmatrix}-0.400459 \\ 0.557354 \\ 0.293274 \\ -0.665567\end{bmatrix}, \\
|B_{2,3}\rangle &= \begin{bmatrix}-0.509916 \\ 0.021673 \\ 0.618488 \\ 0.597485\end{bmatrix}, &
|B_{2,4}\rangle &= \begin{bmatrix}-0.715887 \\ -0.551646 \\ -0.379278 \\ -0.198344\end{bmatrix}.
\end{align*}
\vspace{2cm}
\nocite{*}

\bibliography{apssamp}

\end{document}